\documentclass[11pt]{article}

\usepackage{mathtools}

\usepackage{graphicx}
\usepackage{amssymb}
\usepackage{amsmath}
\usepackage{upgreek}
\usepackage{multirow}
\usepackage[latin1]{inputenc}
\usepackage[labelfont=bf,labelsep=period,footnote size]{caption}
\usepackage{color}
\usepackage[breaklinks=true,colorlinks=true,linkcolor=blue,citecolor=blue,backref=none,pagebackref=false]{hyperref}
\usepackage{setspace}
\usepackage{epstopdf}

\usepackage{xcolor}
\definecolor{c1}{rgb}{0,0,1} 
\definecolor{c2}{rgb}{0.1,0.1,0.1} 
\definecolor{c3}{rgb}{0.3,0,0.9} 
\hypersetup{
    linkcolor= {c1}, 
    citecolor={c2}, 
    urlcolor={c3} 
}

\numberwithin{equation}{section}

\pdfoutput=1
\usepackage{geometry}
\usepackage[auth-sc-lg,affil-sl]{authblk}

\makeatletter
\renewcommand{\@biblabel}[1]{\quad#1.} 
\makeatother

\usepackage{setspace}
\usepackage{fancyhdr}
\fancypagestyle{empt}{
\fancyfoot[C]{\thepage}}
\begin{document}

\title{Applied Antifragility in Natural Systems: Evolutionary Antifragility}

\author[1]{Cristian Axenie}
\author[2]{Roman Bauer}
\author[3]{Oliver L\'opez Corona}
\author[3]{Elvia Ram\'irez-Carrillo} 
\author[4,5]{Ari Barnett}
\author[5]{Jeffrey West}

\affil[1]{Department of Computer Science and Center for Artificial Intelligence, Nuremberg Institute of Technology Georg Simon Ohm, Nuremberg, Germany.}
\affil[2]{Computer Science Research Centre, University of Surrey, Surrey, United Kingdom}
\affil[3]{Investigadores por Me\'xico (IxM) at Instituto de Investigaciones en Matem\'aticas Aplicadas y en Sistemas (IIMAS), Universidad Nacional Aut\'onoma de M\'exico (UNAM), Ciudad Universitaria, CDMX, Mexico.}
\affil[4]{Department of Molecular Biosciences, University of South Florida, Tampa, FL, USA}
\affil[5]{Department of Integrated Mathematical Oncology, H. Lee Moffitt Cancer Center and Research Institute, Tampa, FL, USA}

\label{firstpage}

\maketitle 

\begin{abstract}
This chapter introduces evolutionary antifragility as the time-scale interaction characteristics of a natural dynamic system. It describes the benefit derived from input distribution unevenness, based on the emergent system dynamics and its uncertain and volatile interactions with the operating environment described by unknown disturbances.  We consider methods for the detection, analysis, and modeling of cancer, environmental, microbiota, and social systems antifragility.\footnote{This is a preprint of the following work $:$ Cristian Axenie, Roman Bauer, Oliver L\'opez Corona, Jeffrey West, Applied Antifragility in Natural Systems From Principles to Applications, Springer Nature reproduced with the permissions of Springer Nature. The final authenticated version is available online at \url{https://link.springer.com/book/9783031903908}}\end{abstract}

\newpage

\onehalfspacing

%
%
%
\section{Evolutionary antifragility}
\subsection{Defining evolutionary antifragility}
The previous chapter has illustrated the connection between the convexity of the payoff function, $f(x)$, and the outcome benefit to input variation. Here, we extend this idea to illustrate the connection between convexity and the statistical properties of the distribution of $f(x)$. There are many biological systems for which it is difficult to measure the system's response function. Alternatively, the system's payoff function may indeed be well-characterized, but external signals or noise introduce additional, unpredictable nonlinearities. We will illustrate several examples of observed antifragile behavior in the outcome distributions despite an unknown or unmeasurable response function.


\begin{figure}[t]
\centering
\includegraphics[width=1\linewidth]{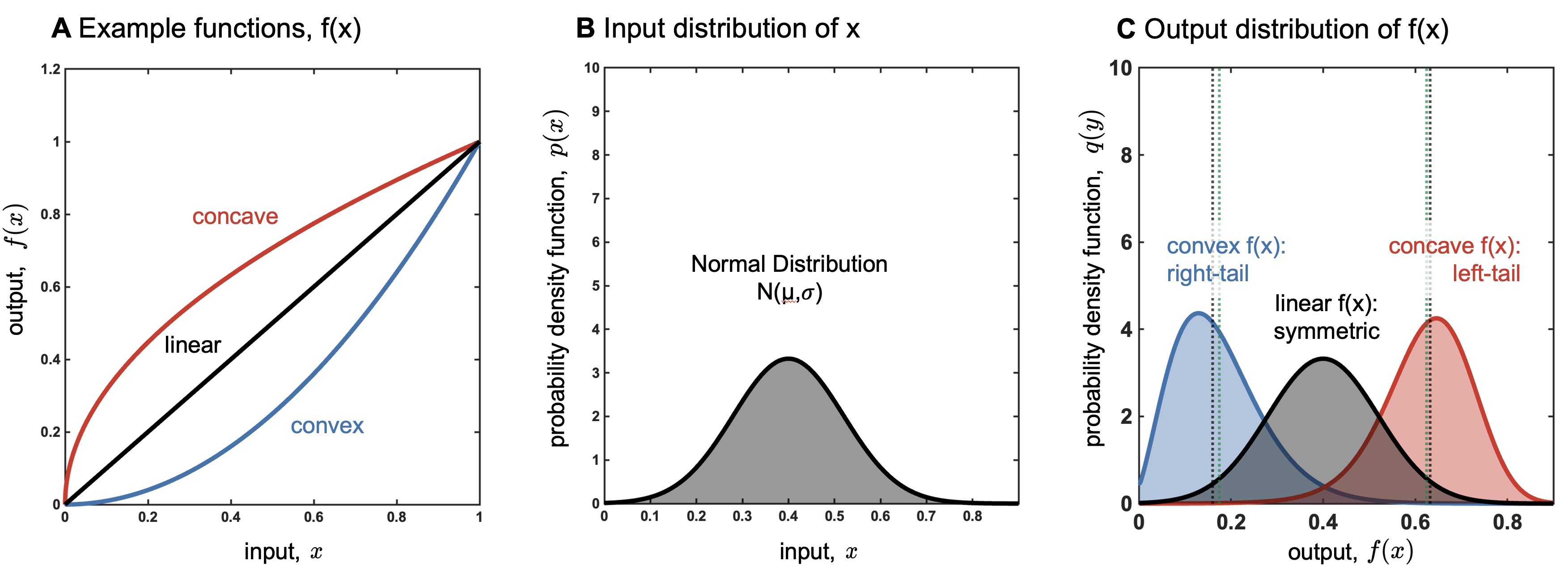}
\caption{(A) Three example functions are shown. Blue: convex function where $f(x) = x^2$. Red: concave function $f(x) = x^{\frac{1}{2}}$. Black: linear function where $f(x) = x$. (B) Consider a continuous random variable, $\mathbf{X}$, which has an associated probability density function that is normally distributed with mean $\mu=0.4$ and standard deviation $\sigma=0.12$. (B, C) We plot the probability density function representing the distribution of outcomes, $f(x)$ for the corresponding convex (blue), concave (red), and linear (black) in A. For convex functions (B), the right tail drives the mean of the outcome distribution up so that $\mathbb{E}(f(x)) > f(\mathbb{E}(x))$. For concave functions, the left tail drives the mean down, so that $\mathbb{E}(f(x)) < f(\mathbb{E}(x))$. For linear functions, the outcome distribution is symmetric (no tail).}
\label{convolution_example}
\end{figure}

\subsection{Probability convolution for convex and concave functions}
We begin with a simple example function and use probability convolution methods to derive the exact expression of this outcome distribution of $f(x)$, given some input probability distribution of $x$. These probability distribution functions (p.d.f.) will have associated properties depending on if the underlying function is convex, concave, or linear.
Consider the following function:
\begin{equation}
    f(x) = x^c
\end{equation}
\noindent where $c$ is a constant. As seen in Fig.~\ref{convolution_example}A, the shape of this function can be convex ($c>1$; blue), concave ($c<1$; red), or linear ($c=1$; black). Let \(p(x)\) be the probability density function describing the input distribution of $x$. For example, Fig.~\ref{convolution_example}B shows a normal distribution with corresponding $\mu$, $\sigma$: $p(x)=N(\mu,\sigma)$:
\begin{equation}
    p(x)=\frac{1}{\sigma \sqrt{2 \pi}} e^{-\frac{1}{2}\left(\frac{x-\mu}{\sigma}\right)^2}.\label{pdf_PFS}
\end{equation}
The probability density function of the outcome, $f(x)$, can be determined by convolution. Let \(X\) be a random variable (input) and let $y(X) = f(X)$, then the outcome distribution of $f(x)$ is given by:
\begin{eqnarray}
P(y(a) \leq Y < y(b) ) &=& \int_a^b p(x) dx \\
 &=& \int_{y(a)}^{y(b)} p(x(y)) \left | \frac{dx}{dy} \right | dy
\end{eqnarray}
where the argument inside the integral is the p.d.f. of $f(x)$, which we denote as $q(y)$:
\begin{equation}
    q(y) = p(x(y)) \left | \frac{dx}{dy} \right |.
\end{equation}
For the example function $f(x)$ given above, both terms of the integrand can be found analytically:
\begin{equation}
    x(y) = y^{\frac{1}{c}},
\end{equation}
and
\begin{equation}
\left|\frac{dx}{dy} \right| = \frac{1}{c} y^{\frac{1-c}{c}},
\end{equation}
which gives:
\begin{equation}
    q(y) = \frac{ y^{\frac{1-c}{c}} }{\sigma c \sqrt{2 \pi}} e^{-\frac{1}{2}\left(\frac{y^{\frac{1}{c}}-\mu}{\sigma}\right)^2}. \label{cdf_PFS}
\end{equation}

The key parameter is $c$, which determines the tailedness of the outcome distribution, as shown in Fig.~\ref{convolution_example}C. Convex functions ($c>1$) are associated with right tails, which cause an increase in the mean of the distribution, such that $\mathbb{E}(f(x)) > f(\mathbb{E}(x))$. Concave functions ($c<1$) are associated with a left-tail, leading to the opposite conclusion: $\mathbb{E}(f(x)) < f(\mathbb{E}(x))$. The implications for systems where the system response function $f(x)$ is unknown are important. Because of the nature of probability distributions, everything fragile (left-tailed) must be concave, while everything antifragile (right-tailed) must be convex \cite{axenie2024antifragility}. Thus, it's possible to determine the benefit (or harm) from input variation through knowledge of the tailedness of the outcome distribution, without knowledge of the underlying functional response, $f(x)$. In the following sections, we illustrate an example of this approach in medicine.

\subsection{Evolutionary antifragility in medicine}
\subsubsection{Inferring (anti)-fragility from Kaplan-Meier curves}
\begin{figure}
\centering
\includegraphics[width=1\linewidth]{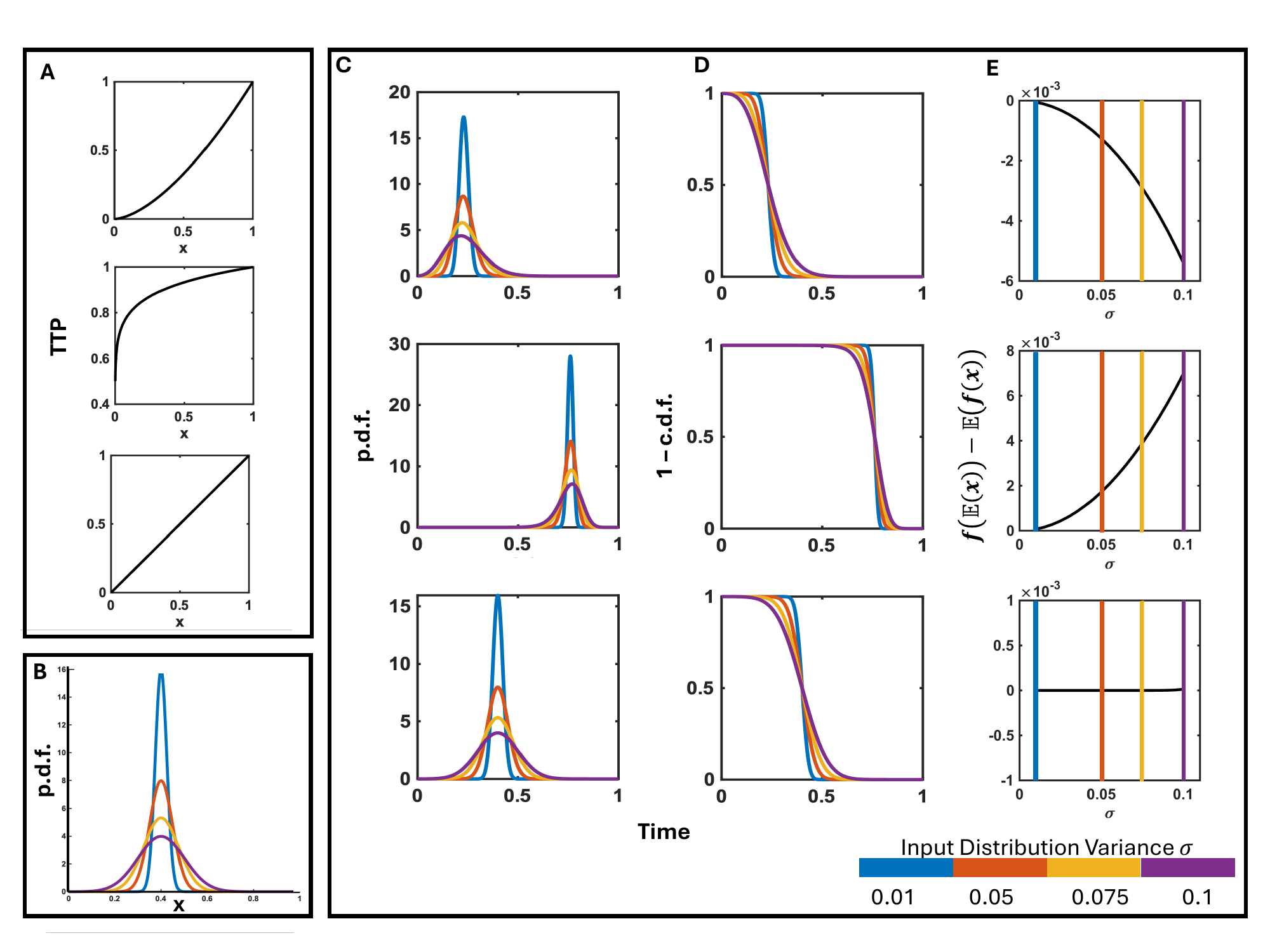}
\caption{(A) Input response function, for an input function $f(x)=x^c$, the convexity of the result can be determined via manipulation of c. The result is one of three potential outcomes: convex, concave, or linear. (B) Dose distribution p.d.f., Given the previous input function, we can state that the mean dose administered is $\mathbb{E}(x)=0.4$. Simply by changing the observed standard deviation ($\sigma$) of the p.d.f., we can see a variety of potential for the dose received. (C) Outcome distribution p.d.f. and resulting Kaplan Meier (D), this can be expanded to observations within the tailedness of the outcome distribution p.d.f. When input functions are convex, the resulting outcome distribution will be right-tailed. Conversely, the opposite will be seen when input functions are concave. (E) Comparison of Jensen's gap. Within nonlinear functions, due to the convexity/concavity of the function: $f(\mathbb{E}(x)) \geq \mathbb{E}(f(x))$ or $f(\mathbb{E}(x)) \leq \mathbb{E}(f(x))$ respectively, will be true due to Jensen's inequality. In linear functions, $\mathbb{E}(f(x))$ and $f(\mathbb{E}(x))$ are identical, and thus, Jensen's gap is zero for all values of $\sigma$. }
\label{mean_example}
\end{figure}

Kaplan-Meier (KM) curves are ubiquitous in clinical trials to determine the efficacy of novel drugs or treatment protocols. KM curves show the fraction of patients who have not yet experienced disease progression, as a function of time~\cite{bland1998survival}. In this example, we model a single patient's time to progression (TTP) and illustrate the differences in KM curves across an entire patient cohort when the underlying functional relationship between dose sensitivity and TTP is either convex, concave, or linear (see Fig.~\ref{mean_example}).

Let $X$ be a random variable (dose sensitivity, $x$) and let the $y(X) = f(X)$ represent the time to progression as a function of dose sensitivity:
\begin{equation}
    f(x) = x^c.
\end{equation}
In the previous section, we have performed the analytical solution of the convolution of the system response function, $f(x)$, and the input p.d.f., $p(x)$  (see eqn.~\eqref{pdf_PFS}) and its corresponding c.d.f (see eqn.~\eqref{cdf_PFS}). Consider the three types of functions in Fig.~\ref{mean_example}A (convex, concave, linear), with associated input variance, $p(x)$ (Fig.~\ref{mean_example}B). Here, the variance in dose, $x$, represents the inter-patient differences in drug delivery, tumor sensitivity, and other sources of patient-to-patient heterogeneity in drug response. We show the resulting p.d.f in Fig.~\ref{mean_example}C. By integrating the p.d.f., we can plot the KM curve (equivalent to $1 - $ c.d.f. with no censoring) shown in Fig.~\ref{mean_example}D. This approach is similar to previous frameworks that primarily utilize numerical methods to convolve inter- and inter-patient variation in drug sensitivity to fit KM curves~\cite{Hwangbo2023, patterson2024ultrasensitive}. The advantage of the approach shown here is that the analytical relationship between the input distribution, $p(x)$ and output distribution, $q(y)$, is known and does not require numerical approximation.

While KM curves are considered the gold standard of medical trial reporting, information can often visually obscure the tailedness of the underlying distribution. In Fig.~\ref{mean_example}D, increasing the value of $\sigma$ does not shift the inflection point (the time at which 50\% of patients have progressed), but there are differences before and after this inflection. These are better illustrated in the corresponding p.d.f (Fig.~\ref{mean_example}C), where the resulting tailedness is noticeably different. As expected, convex functions are right-tailed, while concave functions are left-tailed, and linear functions are perfectly symmetric about the mean.

In a treatment context, this can be explained through the importance of extreme dose sensitivity (both high and low) and its effect on outcome. For a given population's drug sensitivity (mean and variance), if the therapeutic benefit has decreasing returns as dose increases as in a concave response, extreme doses would contribute overall less to the mean value of the therapeutic outcome. Conversely, convex dose response outlier patients will amplify the population's mean responsiveness, especially when input variance is larger. The difference in the mean values (e.g. the antifragility) can be quantified as follows: 
\begin{eqnarray}
 f(\mathbb{E}(x)) - \mathbb{E}(f(x)) &=& \mu^c - \int_{-\infty}^{\infty} y q(y) dy 
\end{eqnarray}
The differences are shown in Fig.~\ref{mean_example}E, confirming that this difference is positive (antifragile) for convex functions, negative (fragile) for concave functions and zero for linear functions. Therefore, by understanding the underlying distribution function and its variability, the tradeoff between efficacy and safety, as well as risk, can be better assessed. This would then allow for a better understanding of how variability impacts the desired clinical outcome. It can be used to inform dosing strategies for personalized medicine applications. When the underlying input distribution is known (or can be estimated), the resulting p.d.f. of outcomes can be obtained from just a KM curve via backward convolution.


\subsubsection{Current applications in medicine}
These concepts can directly tie to the clinical practice of adaptive therapy, in which the prescribed dosing of treatment is continuously adjusted based upon tumor evolution and disease progression~\cite{gatenby2009adaptive}. Adaptive therapy is not used as an extinction therapeutic but more commonly as a way for disease management/control while maintaining a high degree of quality of life \cite{cunningham2018optimal, hansen2020cancer}. Dose modulations are designed to promote cell-cell competition among sensitive and resistant cell populations~\cite{gatenby2009adaptive, west2018capitalizing}. The key difference between adaptive therapy and standard-of-care approaches is the increased variance in dose delivered~\cite{west2024fundamentals}. The benefits arising from input perturbations illustrate how adaptive therapy's underlying philosophy is related to antifragility. By allowing for perturbations, patient response can be seen to be increasingly positive while maintaining other clinically important factors (e.g. limiting resistance emergence). 

Many conditions outside of cancer can likewise be improved through perturbations to the environment. One such example is the application of radiotherapy in patients with Osteoarthritis (OA). Characterized by a decrease in mobility and increasing pain and stiffness of joints, current treatment options for OA range from lifestyle modification to total joint arthroplasty. Radiotherapy provides an alternative, minimally invasive option with proven success where low-dose application was used (a dose of 0.5Gy per fraction for 6 fractions given every other day or twice weekly) \cite{Dove2022,mucke2018leitlinen,alvarez2021radiotherapy}. These small environmental perturbations allow for multiple mechanisms of anti-inflammatory effects, such as an increase in anti-inflammatory cytokines (e.g. IL10, TGF-$\beta$1), decreased production of reactive oxygen species, and increased polarization of Macrophages to the M2-phenotype \cite{HILDEBRANDTMPSEEDCNFREEMA1998}. 

The benefit to the patient lies not only in the therapies' mechanism of action but also its schedule. Previous work applied antifragile theory to measure second-order effects on resistance, collateral sensitivity, and combination treatments \cite{west2020antifragile}. In the context of resistance, there is an expanded range of dose values classified as antifragile after the evolution of resistance has occurred \cite{west2020antifragile}. This would indicate that treatment schedules with increased perturbations are more beneficial. This effect is exaggerated when accounting for pharmacokinetics and drug delivery \cite{pierik2024second}. For collateral sensitivity, it is possible to maintain secondary vulnerabilities to a subsequent therapy line. This is supported in part by the shifting of cell lines from fragile to antifragile responses during treatment.





\newpage



\end{document}